\documentclass[12pt,preprint]{aastex}

\usepackage{graphicx}
\usepackage{dcolumn}
\usepackage{bm}

\renewcommand{\vec}[1]{\boldsymbol{#1}}

\newcommand{\Alfven}{Alfv\'{e}n }

\begin{document}

\shorttitle{Relativistic Current Sheet}
\shortauthors{Zenitani \& Hoshino}

\title{Relativistic Particle Acceleration in a Folded Current Sheet}

\author{S. Zenitani and M. Hoshino}
\affil{
Department of Earth and Planetary Science, University of Tokyo,
7-3-1, Hongo, Bunkyo, Tokyo, 113-0033 Japan;
zenitani@eps.s.u-tokyo.ac.jp
}

\begin{abstract}
Two-dimensional particle simulations of
a relativistic Harris current sheet of
pair ($e^{\pm}$) plasmas
have demonstrated
that the system is unstable to the relativistic drift kink instability
and that a new kind of acceleration process takes place
in the deformed current sheet. 
This process contributes to the generation of non-thermal particles and
contributes to the fast magnetic dissipation in the current sheet structure.
The acceleration mechanism and
a brief comparison with relativistic magnetic reconnection are presented.
\end{abstract}


\section{Introduction}
A current sheet structure with reversed magnetic fields
can be seen everywhere in the universe.
Among the magnetic field dissipation processes in the current sheet,
magnetic reconnection may be the most popular one.
Breaking the topology of magnetic field lines,
it releases the magnetic energy into particles' kinetic energy
in a relatively short time-scale
(several to several tens of an \Alfven transit time $\tau_{A} = \lambda/V_{A}$,
where $\lambda$ is the half thickness of the current sheet and
$V_{A}$ is the typical \Alfven speed).
It plays an essential role in solar terrestrial sites
such as the Earth's magnetotail \citep{dungey}
and source regions of the solar flare \citep{parker}.
Moreover, it is expected in astrophysical sites such as
active galactic nuclei \citep{les98,schopper} and pulsar winds,
especially their well-studied example of the Crab Nebula \citep{coro90}.
In case of the Crab Nebula,
energy conversion in a relativistic wind of $e^{\pm}$ plasmas has been
a long-standing problem (the so-called the ``$\sigma$-problem'',
where $\sigma$ is the ratio of the Poynting flux energy to the particle kinetic flux)
\citep{kc84,coro90,lyu01,kirk03}.
The wind is originally Poynting-dominated ($\sigma \sim 10^4$) close to the neutron star
\citep{arons74};
but it is kinetic-dominated near the termination shock
($\sigma$ is far less than the unity;
$\sigma \sim 10^{-3}$ by \citet{kc84} and $\sigma \sim 10^{-1}$ by \citet{mori04}).
Recently,
relativistic magnetic reconnection in the striped current sheets
has been discussed as the most possible dissipation mechanism
\citep{coro90,lyu01,kirk03}.
The properties of relativistic reconnection in $e^{\pm}$ plasmas
have been investigated
by an analytical study \citep{blackman} and
by computer simulations \citep{zeni01,claus04}.
One of the most important features of relativistic reconnection
is the generation of non-thermal particles
due to direct acceleration by dc electric field around the $X$-type region \citep{zeni01}.
However, the above discussion is based on
a two-dimensional picture of reconnection.
Instabilities in a cross-field plane are also of critical importance
to understanding a realistic current sheet problem that includes reconnection.
Does reconnection grow faster than cross-field instabilities?
Is reconnection the fastest dissipation process?
In the case of non-relativistic ion-electron plasmas,
the following instabilities have been discussed:
the lower hybrid drift instability (LHDI) \citep{krall71,dav77},
the Kelvin-Helmholtz instability (KHI) \citep{yoon96,shino01},
and the drift kink instability (DKI) \citep{zhu96,prit96,dau98}.
They lead to plasma heating and particle acceleration
as well as the triggering of reconnection.
The purpose of this Letter is to search for the basic properties
of such cross-field instabilities and of the non-thermal particle generation
in a relativistic $e^{\pm}$ current sheet.
In this Letter, we report our 2D simulations in a cross-field plane
of a relativistic $e^{\pm}$ current sheet and
we report a new acceleration process related to
the relativistic drift kink instability (RDKI) \citep{icrc03}.

\section{Simulation}

The simulation code used here is a three-dimensional particle-in-cell code.
The system size is composed of 1 ($X$) $\times$ 256 ($Y$) $\times$ 512 ($Z$) grids.
We consider periodic boundaries in the $X$, $Y$ directions,
and we periodically set two simulation boxes in the $Z$ direction.
The typical scale of the current sheet $\lambda$ is set to 10 grids,
so that the Y boundaries are located at $y = \pm 12.8 \lambda$,
and the Z boundaries are located at $z = \pm 12.8 \lambda$.
We take the relativistic Harris model
as the initial current sheet.
The magnetic field, the density of plasmas and
their distribution functions are described by
$\bm{B} = B_0 \tanh(z/\lambda) \hat{\bm{x}}$,
$n(z) = n_0 \cosh^{-2} (z/\lambda)$ and
$f_{\pm} \propto n(z) \exp[-\Gamma_\beta\{\varepsilon - \beta_\pm u_y\} / T ] $,
respectively.
In the above equations,
$B_0$ is the magnitude of magnetic field in the lobe (background region),
$n_0$ is the number density of plasmas in the current sheet,
$\beta_\pm = v_{\pm}/c$ are the drift velocities for each species;
$\beta_{+} = + \beta$ for positrons,  $\beta_{-} = - \beta$ for electrons,
$\Gamma_\beta$ is the Lorentz factor for $\beta$ [$\Gamma_\beta = [1-\beta^2]^{-1/2}$],
$\varepsilon$ is the particle energy,
$\bm{u}$ is the relativistic four velocity of $\bm{u}= [1-(v/c)^2]^{-1/2} \cdot \bm{v}$
and $T$ is the temperature.
In the first simulation, we set $T = mc^2$ and $\beta = 0.3$.
We set no driving force to excite instabilities,
so that instabilities arise from thermal noise.
We set no background plasmas.
The total energy is conserved within an error of $0.3\%$
throughout the simulation run.

\section{Results}

\begin{figure*}
\begin{center}
\includegraphics[width={\columnwidth},clip]{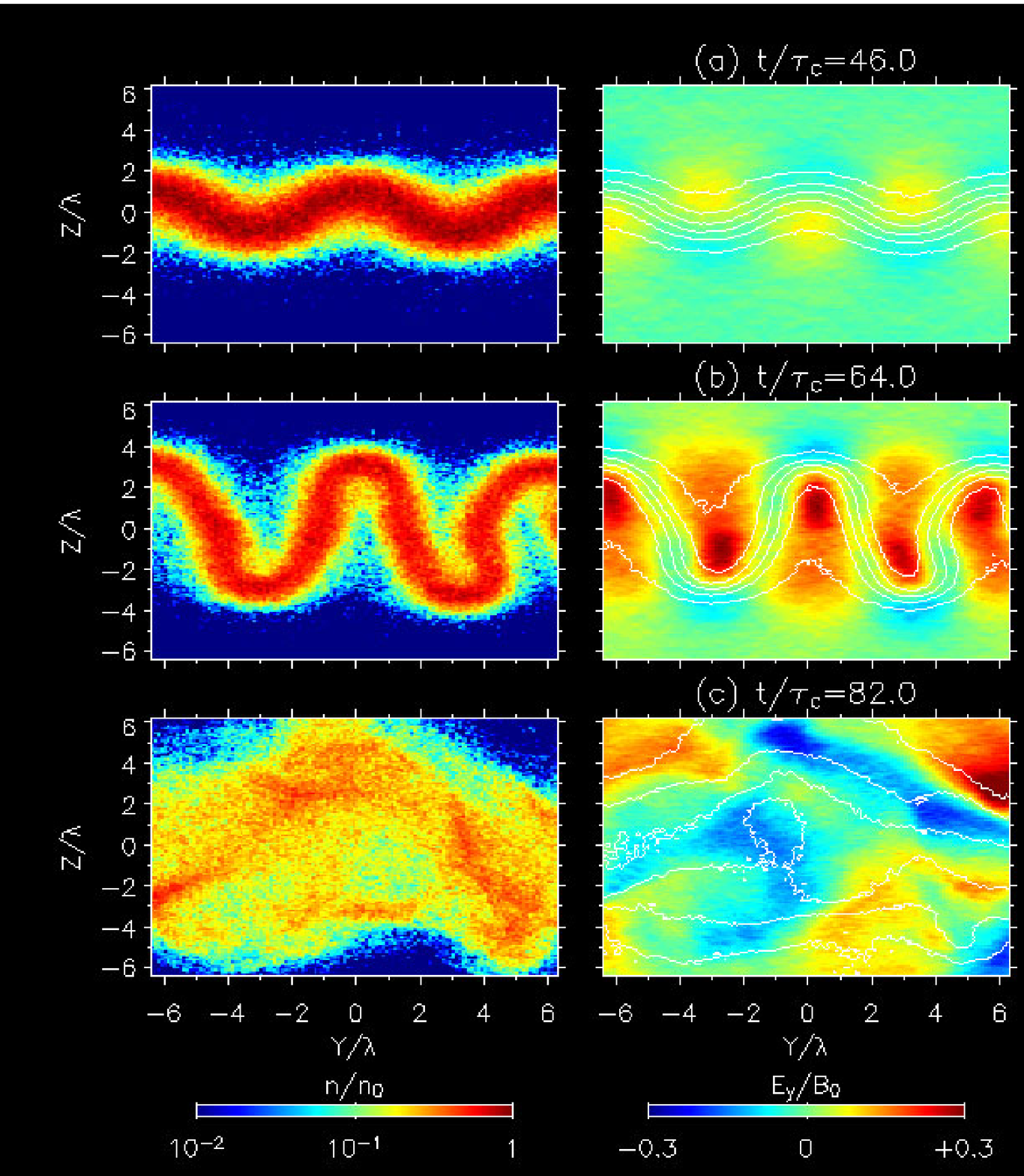}
\caption{\label{fig:kink}
Left three panels: Color contours of
the plasma density at $t/\tau_c=$46.0, 64.0 and 82.0.
Middle three panels: $E_y$ at the corresponding time;
the white lines shows contours of $B_x$.
Right three panels: Energy spectra.
}
\end{center}
\end{figure*}

Snapshots at three characteristic stages of the simulation
are presented in Figure~\ref{fig:kink}.
The three left panels in Figure~\ref{fig:kink} show color contours of
the plasma density at
($a$) $t/\tau_{c}=46.0$, ($b$) $64.0$ and ($c$) $82.0$,
where time is normalized by the light transit time $\tau_{c}=\lambda/c$.
The three middle panels show
color contours of the electric field $E_y$ at the corresponding time;
the white lines show contours of $B_x$.
The three panels show the
particles' energy spectra for the whole simulation.
The horizontal axis shows the particles' energy normalized by $mc^2$.

As shown in the top three panels in Fig.~\ref{fig:kink},
we observe a kink-type modulation of the current sheet.
We confirmed that this is a signature of
the RDKI,
a relativistic extension of the DKI,
which is a long-wavelength, current-driven instability in a thin current sheet.
In the top middle panel, one can see the typical structure of
$E_y$ components of the polarization of electric fields of RDKI.
The sign of $E_y$ is positive
in the yellow regions,
while
it is negative in the blue regions.
Two types of regions appear alternatively along the current sheet 
and they are anti-symmetric with the neutral plane ($z=0$).
Note that the $\vec{E} \times \vec{B}$ direction using
the lobe magnetic fields and the observed sign of $E_y$
is consistent with the $Z$-displacement of the plasma bulk motion.
The symmetric $E_z$ components are also observed and
their values are a little less than or comparable to those of $E_y$.
The $E_x$ components are negligible.
Since the system is uniform in the $X$ direction,
there is no electrostatic $E_x$ fields.
After $t/\tau_{c} \sim 50$, the instability turns into its non linear stage.
The three middle panels in Fig.~\ref{fig:kink} are
snapshots of the current sheet at $t/\tau_{c} = 64.0$.
The current sheet is strongly folded 
within a zone of $-4 < z/\lambda < 4$.
Due to the $Z$-displacement by RDKI,
red (originally yellow) regions, where $E_y>0$, are placed in a row
around the neutral plane ($z = 0$)
in the midmost panel of Fig.~\ref{fig:kink}.
The total kinetic energy in the system has increased by 30\% at this time.
Importantly, we observe a clear sign of the particle acceleration
in a high-energy tail in the energy spectrum of the middle-right panel in Fig.~\ref{fig:kink}.
We will discuss the acceleration process later.
The kinked-shape of the current sheet is not stable,
because fragments of the anti-parallel  currents ($\pm J_z$) pull each other.
As they start to collide, the system turns into the ``mixed'' stage.
The snapshots at this stage are presented
in the bottom three panels in Fig.~\ref{fig:kink} at $t/\tau_{c} = 82.0$.
Due to the enhanced diffusion by sheet collisions,
the total kinetic energy has increased by 170\% at this time.
In the energy spectrum of the bottom-right panel,
one can recognize
both a remnant of the non-thermal tail and global plasma heating.
Finally, the mixed current sheet slowly evolves into the broadened current sheet,
which is 3-4 times thicker than the initial state.
The saturated level of the total kinetic energy is 330\%-340\% at $t/\tau_{c} = 200$ or later.
The spectrum looks unchanged from the mixed stage; one can still observe the non thermal tail.

\section{Discussion}

\begin{figure}
\begin{center}
\includegraphics[width={0.9\columnwidth},clip]{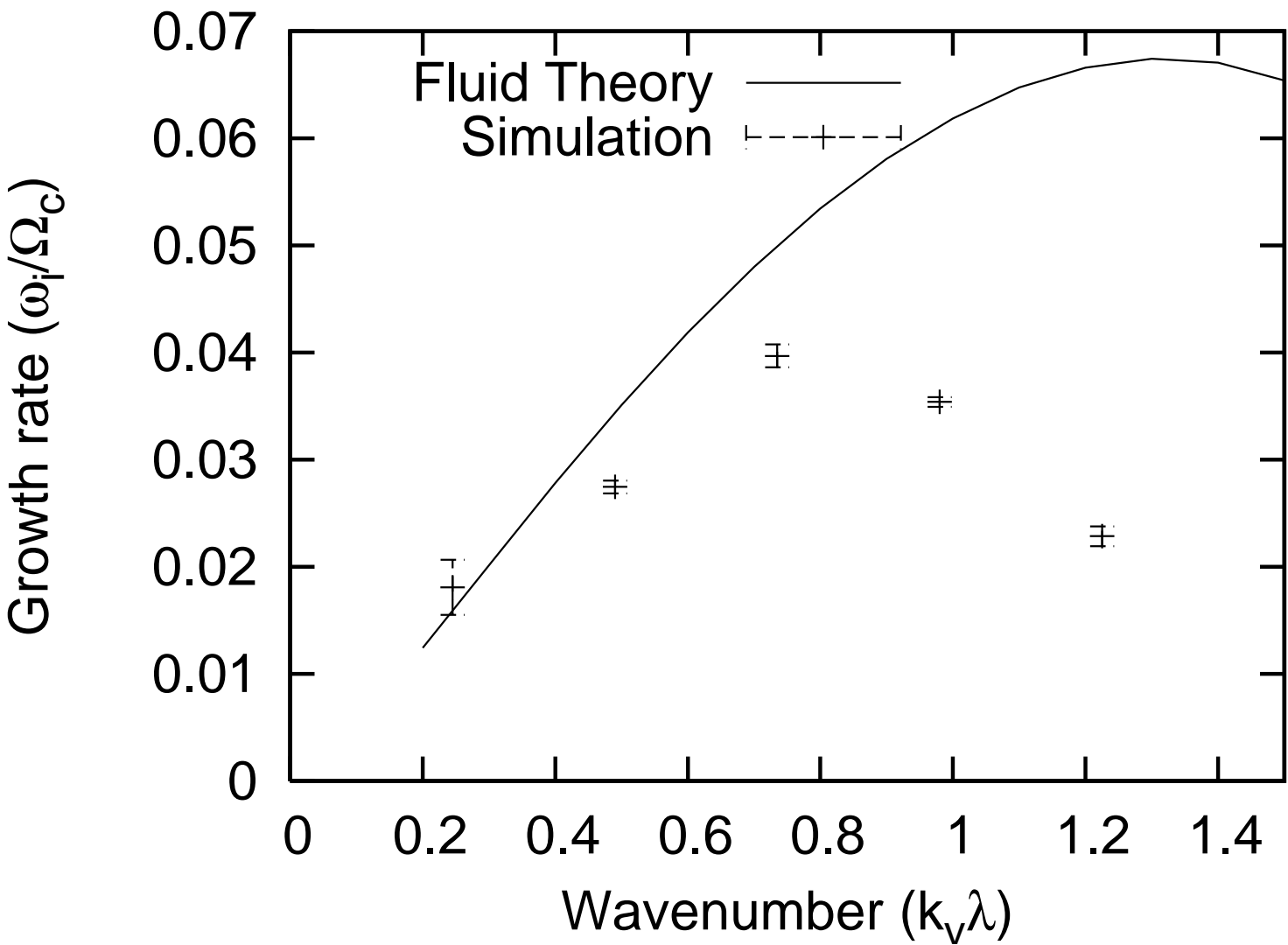}
\caption{Plot of the observed growth rates ($\omega_i / \Omega_c$)
as a function of the normalized wavenumber ($k_y \lambda$).
The solid line shows the growth rate predicted by relativistic two-fluid theory.
\label{fig:growth}}
\end{center}
\end{figure}

Now, we again take a look at the instability during its linear stage.
The observed growth rate ($\omega_i/\Omega_c$) is plotted
as a function of $k_y \lambda$ in Fig.~\ref{fig:growth},
where $k_y$ is the wavenumber
and $\Omega_c$ is the gyro frequency $\Omega_c=\omega_c/\gamma=(eB)/(\gamma mc^2)$.
The growth rate and wave length
of the most dominant mode of RDKI
are $\omega_i/\Omega_c \sim 0.035$ and $k_{y}\lambda \sim 1.0 $, respectively.
We have also calculated eigen functions and their growth rates of the RDKI,
by linearizing relativistic two-fluid equations.
The theoretical growth rates for the present case are shown as a solid line in Fig.~\ref{fig:growth}.
The simulation results are consistent with them
except for shorter wavelength of $k_y \lambda > 1$,
where the instability is suppressed by kinetic effects.
We observe a slight signature of the relativistic drift sausage instability (RDSI) along with RDKI.
However, RDSI is less influential on the current sheet deformation and particle acceleration.
Note that no velocity-shear-driven mode such as KHI  is excited,
ant that is because there is no velocity shear between
drifting plasmas in the current sheet
and empty background plasmas.
Also note that LHDI is not excited in an $e^{\pm}$ case,
because there is no delay of a positron's response to an electron's.

\begin{figure}
\includegraphics[width={0.95\columnwidth},clip]{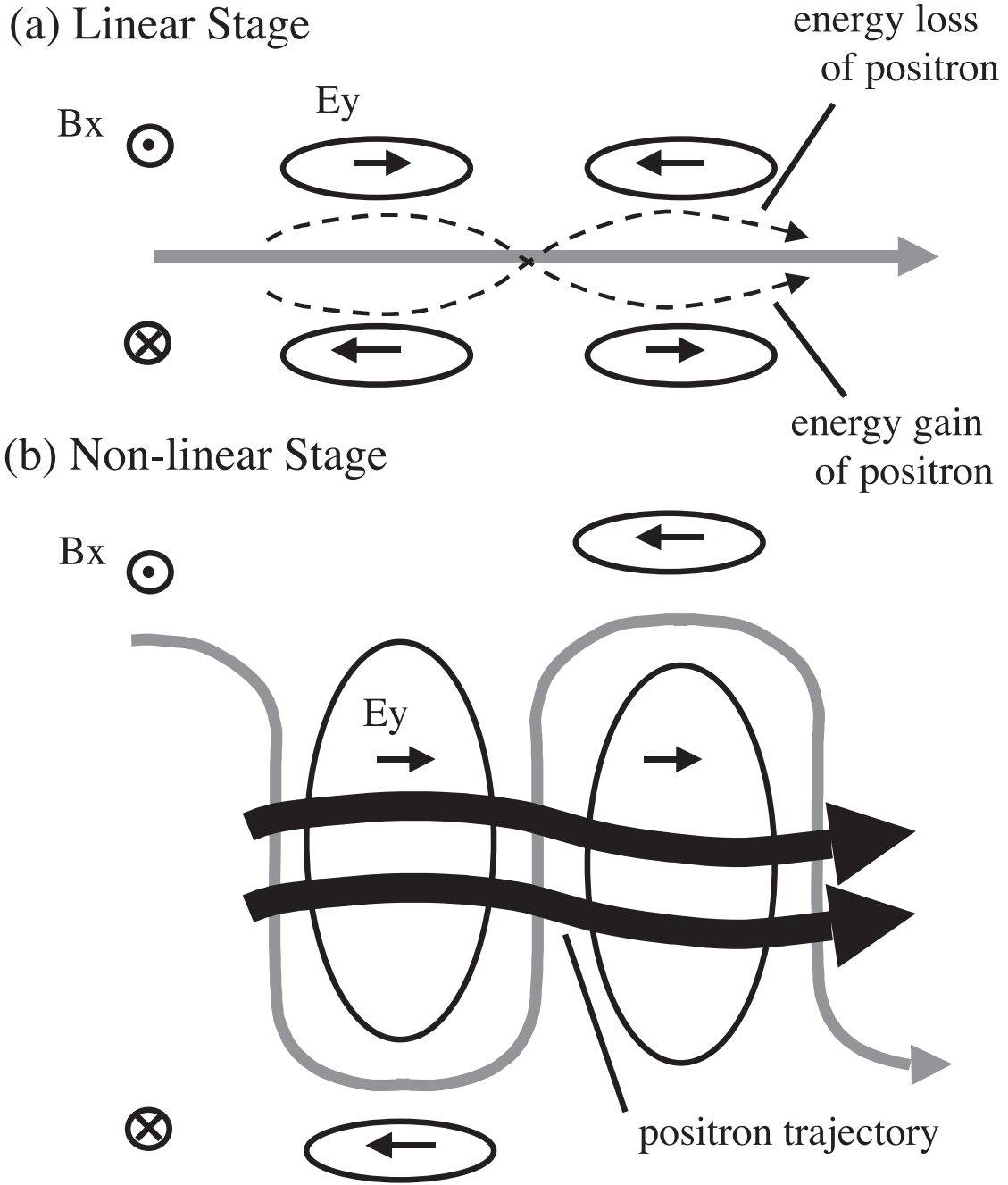}
\caption{\label{fig:accel}
Schematic view of the acceleration mechanism.
Once the current sheet is folded at the non linear stage,
regions with the positive $E_y$ sign are located in a central acceleration channel.}
\end{figure}

Next,
we discuss an acceleration mechanism that is found at the non linear stage of RDKI.
By analyzing trajectories of highly accelerated particles,
we found that the main site of particle acceleration is
the central channel around $z \sim 0$,
where the current sheet was originally located.
For simple discussion, we call this channel the ``acceleration channel (AC)''.
We illustrate the acceleration mechanism
in schematic views in Fig.~\ref{fig:accel}.
In the upper view,
the structure of the polarization of $E_y$ fields at the linear stage are illustrated.
They are anti symmetric with respect to $Z$
and both positive and negative regions are alternatively located along the sheet.
Passing through their meandering orbits,
some particles can resonate with the $E_y$ fields.
Some of them gain their energy and others lose their energy.
At the non-linear stage, due to the $Z$-displacement of the current sheet,
the positive $E_y$ regions are now located around the AC.
nside the positive $E_y$ regions, typical values of the fields are
$|B_x| \approx 0.6-0.7 B_0$ and $E_y \approx 0.2-0.3 B_0$.
Contrary to the reconnection \citep{zeni01,claus04}
in which $|E|/|B|>1$ in an acceleration site,
the value of $|E|/|B|$ is less than unity 
inside the positive $E_y$ regions,
and so
low-energy particles travel into the $E \times B$ direction.
However, fewer populations of high-energy particles
can travel across the positive $E_y$ regions,
because their Larmor radii $r_L$ are larger than
the half-wavelength of RDKI ($\pi/k_y$).
They gain their energy from the electric field $E_y$,
cross the current sheet and move into the neighboring positive $E_y$ regions.
Since $r_L$ becomes larger as particles gain more energy
($r_L \propto \varepsilon^{1/2}$ in non-relativistic case and
$r_L \propto \varepsilon$ in relativistic case),
accelerated particles usually satisfies the crossing condition of $r_L>\pi/k_y$
in the following positive $E_y$ regions.
In this way, they are successively accelerated in the AC,
passing through multiple positive $E_y$ regions into the $\pm Y$ direction.
The above condition can be interpreted as the ``trapping condition'' inside the AC.
Higher accelerated particles tend to stay in the AC,
because larger $r_L$ prevents them from escaping.
This is why the non-thermal tail is enhanced in the energy spectrum.
We think this type of enhanced acceleration can be found
in any kink-type instability in a thin current sheet,
as long as its wavelength is comparable to a particle's Larmor radius.
For example, it may be applicable to acceleration of ions,
when a thin current sheet of ion-electron plasmas is modulated by KHI.
The folded structure evolves into the mixed stage
when its $Z$-displacement ($\Delta z$) is nearly the same as the half-wavelength of RDKI
($\Delta z \sim \pi/k_y\sim\pi\lambda$).
Assuming that $v_z$ is typically $V_{E\times B} \sim 0.3c$ in the positive $E_y$ regions,
the time scale of the non-linear stage ($\tau_N$) is approximated by 
$\tau_{N} \sim \pi\lambda/v_z \sim 5-10 \tau_c$.

After the system turns into the mixed stage,
a few number of high-energy particles and
a lot of low-energy particles in the current sheet
interact with the positive $E_y$ regions.
Then the magnetic fields are dissipated in a time-scale of tens of $\tau_c$.
Roughly speaking,
the magnetic field energy is consumed by
Joule heating of the $E_y$ fields
and the total current $\bar{I_y}$ in the broadened current sheet.
The magnetic energy stored between the folded curves is
approximated by $(\pi/k_y)  {B_0^2}/{8\pi} $. 
We assume that the average value of the electric fields
$\bar{E_y} \sim 0.5 E_y = 0.5 \times (0.2-0.3) B_0$, where
the factor 0.5 represents the fact that positive $E_y$ is not uniform in the AC.
The total current $\bar{I_y}$ is represented by the zeroth order current
$\bar{I_y} \sim \lambda J_0  =  c B_0 /4\pi$. 
Thus the dissipation time scale ($\tau_D$) is
approximated by 
$\tau_D  \sim  (\pi/k_y) ( {B_0^2}/{8\pi} )  ( \lambda J_0 \bar{E_y} )^{-1}
\sim
10-20 \tau_c$.

Next, we briefly compare RDKI and magnetic reconnection \citep{zeni01}.
In Table~\ref{table}, the parameters for several simulation runs of RDKI and
for relevant runs of reconnection are presented.
Throughout the simulation runs, we calculate
the ratio of the non-thermal kinetic energy to the total kinetic energy ($K_{nonth}/K$)
and their maximum values in time are presented in Table~\ref{table}.
In case of run D3, $K_{nonth}$ is not resolved by our method.
The maximum energies ($\varepsilon_{max}$) are also presented in Table~\ref{table}.
In the case of RDKI, the highest energy particles run in the AC into the $\pm Y$ direction
with the light speed
during the non-thermal stage and the mixed stage.
Their maximum energies are estimated by 
$\varepsilon_{est} \sim \varepsilon_{max0} +  ec\bar{E_y} (\tau_N + \tau_D)
\sim
\varepsilon_{max0} + 3 \omega_c \tau_c mc^2$
and
the estimated values agree with the simulation results.
In three cases of reconnection,
the evolution seems to be restricted by the periodic system size of $L_x$,
because two outflow jets travel into the $\pm X$ direction.
Due to this restriction,
the upper limits of particle energy 
are estimated by $\varepsilon_{est}=eB_0L_x/2$
in Table~\ref{table}
in reconnection cases.
However, 
the obtained maximum energies ($\varepsilon_{max}$) by reconnection
are far larger than the maximum energies by RDKI.
From the viewpoint of field dissipation (e.g. striped pulsar winds),
RDKI can dissipate more magnetic energy.
The linear growth rates of the energy conversion ($\tau_c d \ln ( \Delta K ) /dt$)
are presented in Table 1.
Generally speaking, RDKI grows \textit{faster}
in a relativistic current sheet ($T/mc^2 \geq 1$) and
it converts $2.5-3.5$ times as much energy as the initial kinetic energy in the current sheet
in a time scale of $\tau_D$.
Moreover, RDKI spontaneously occurs everywhere in the $X$ direction,
while single reconnection requires a wide spatial range in the $X$ direction
in order for the outflow jets to escape.
Therefore, even in runs R3 and D3, in which RDKI grows slower,
reconnection takes twice as long to dissipate the same amount of the magnetic energy as RDKI.
On the contrary,
reconnection is the more favored candidate as the origin of non-thermal particles.
It produces more non-thermal energy of $K_{nonth}$ than RDKI
as a result of the high ratio of $K_{nonth}/K$.

Our initial results of a three-dimensional simulation show that
RDKI occurs almost uniformly in the current sheet.
Also, ~\citet{claus04} have presented
their three-dimensional results of triggered reconnection.
They have referred to the importance of RDKI in the three-dimensional case and
a signature of a two-dimensional-like RDKI is observed outside the reconnection region.
In case of multiple current sheets structure, such as striped pulsar winds,
there are several possibilities of enhanced field dissipation.
If the distance between two currents is less than
the critical distance of $2\lambda_c$,
the two current sheets collide with each other and
90\% or more of the magnetic energy will be dissipated into the particle energy.
In typical cases, $\lambda_c \sim$ 7-8, but
its dependency on $T$ or $\beta$ should be further investigated.
We also note that
the secondary RDKI takes place in the broadened current sheet
in some occasions.


Let us summarize this Letter.
First, we investigate the RDKI in a thin $e^{\pm}$ current sheet.
Second, we find a new acceleration process that uses a folded structure of a current sheet.
The unique point of this process is that
AC electric fields evolve into a DC acceleration channel.
Third, due to the enhanced acceleration, a non-thermal particles are produced.
Fourth, comparison with reconnection
shows that RDKI is more favorable to dissipate the magnetic energy
in the current sheet structure.
We believe that RDKI provides a crucial clue to
the energy dissipation problem in the current sheets in the universe,
including striped pulsar winds.

\begin{acknowledgments}
This work was supported by the facilitates of
JAXA and the Solar-Terrestrial Environment Laboratory, Nagoya University.
\end{acknowledgments}


\clearpage

\clearpage

\begin{table*}
\begin{center}
\begin{tabular}{l|cccccccccc}
Run &
$T$ &
$L_x$ &
$L_y$ &
$L_z$ &
${K_{nonth}}/{K}$ &
$\tau_c d \ln ( \Delta K ) /dt$ &
$\varepsilon_{max}$ &
$\varepsilon_{est}$ &
\\
\hline
D1 &
16 &
1.0 &
25.6&
51.2 &
$4.7 \times 10^{-2}$ &
$2.3 \times 10^{-1}$ &
$7.0 \times 10^{2}$ &
$7.1 \times 10^{2}$ &
\\
D2 &
1 &
1.0 &
25.6&
51.2 &
$7.2 \times 10^{-2}$ &
$2.3 \times 10^{-1}$ &
$4.8 \times 10^{1}$ &
$4.2 \times 10^{1}$ &
\\
D3 &
1/16 &
1.0 &
25.6&
51.2 &
- &
$1.9 \times 10^{-2}$ &
4.0 &
3.0 &
\\
R1 &
16 &
102.4 &
1.0 &
51.2 &
$>2.5 \times 10^{-1}$ &
$6.6 \times 10^{-2}$&
$>3.6 \times 10^{3}$ &
$5.6 \times 10^{3}$ &
\\
R2 &
1 &
102.4 &
1.0 &
51.2 &
$>2.6 \times 10^{-1}$ &
$6.5 \times 10^{-2}$ &
$>1.2 \times 10^{2}$ &
$3.5 \times 10^{2}$ &
 \\
R3 &
1/16 &
102.4 &
1.0 &
51.2 &
$>1.1 \times 10^{-1}$ &
$7.1 \times 10^{-2}$ &
$>5.3$ &
$2.2 \times 10^{1}$ &
\\
\end{tabular}
\caption{\label{table}
Three runs for RDKI (D1-3) and
relevant runs of relativistic reconnection (R1-3) are presented.
The temperature ($T/mc^2$), the system size ($L_x, L_y, L_z$ in unit of $\lambda$),
the maximum non-thermal ratio of the kinetic energy ($K_{nonth}/K$),
the energy conversion rate ($\tau_c d \ln ( \Delta K ) /dt $),
the maximum energy ($\varepsilon_{max} /mc^2)$
and
their estimated values ($\varepsilon_{est} /mc^2)$
are presented.
}
\end{center}
\end{table*}

\end{document}